\documentclass[10pt,letterpaper,twocolumn]{article} 

\usepackage{ol2}
\usepackage[draft]{hyperref}
\usepackage{amsmath}

\begin{document}

\twocolumn[ 

\title{Improved measurement of polarization state in terahertz polarization spectroscopy}


\author{M. Neshat$^*$ and N. P. Armitage}

\address{Department of Physics and Astronomy, The Johns Hopkins University, Baltimore, Maryland 21218, USA\\
$^*$Corresponding author: mneshat@jhu.edu
}

\begin{abstract}
A calibration scheme is presented for improved polarization state measurement of terahertz pulses. In this scheme the polarization response of a two-contact terahertz photoconductive detector is accurately measured, and is used to correct for the impact of the non-idealities of the detector. Experimental results show excellent sub-degree angular accuracy and at least $60\%$ error reduction with this scheme.
\end{abstract}


 ] 

\noindent 
Terahertz polarimetry is a quickly growing characterization tool for the study of effects such as birefringence and the magneto-optic Kerr effect \cite{Castro-Camus_Nov2011}-\cite{Aguilar_Sep2011}. Recently a number of polarization sensitive methods for terahertz pulse radiation have been proposed \cite{Castro-Camus_May2011}\nocite{Hussain_May2008}\nocite{Makabe_Sep2007}\nocite{Dong_Oct2010}--\cite{Byrne_Mar2011}. Most of these methods are based on photoconductive antenna detectors, in which it is assumed that the detector has an ideal linear polarization response over the entire frequency range. However, it has been shown recently that this assumption is generally not valid \cite{Gong_Sep2011} and that in addition to the geometrical structure, the polarization response of the photoconductive detector depends strongly on the optical and terahertz alignments \cite{Gong_Sep2011,Castro-Camus_May2007}. Therefore, it is clear that calibration schemes are necessary to compensate for a non-ideal polarization response of the detector in order to get the highest accuracy.  In this letter, we propose a calibration scheme for widely used two-contact photoconductive detectors for accurate measurement of the polarization state of pulsed THz radiation.
  
\begin{figure*}[!hbtp]
\centerline{\includegraphics[width=10cm]{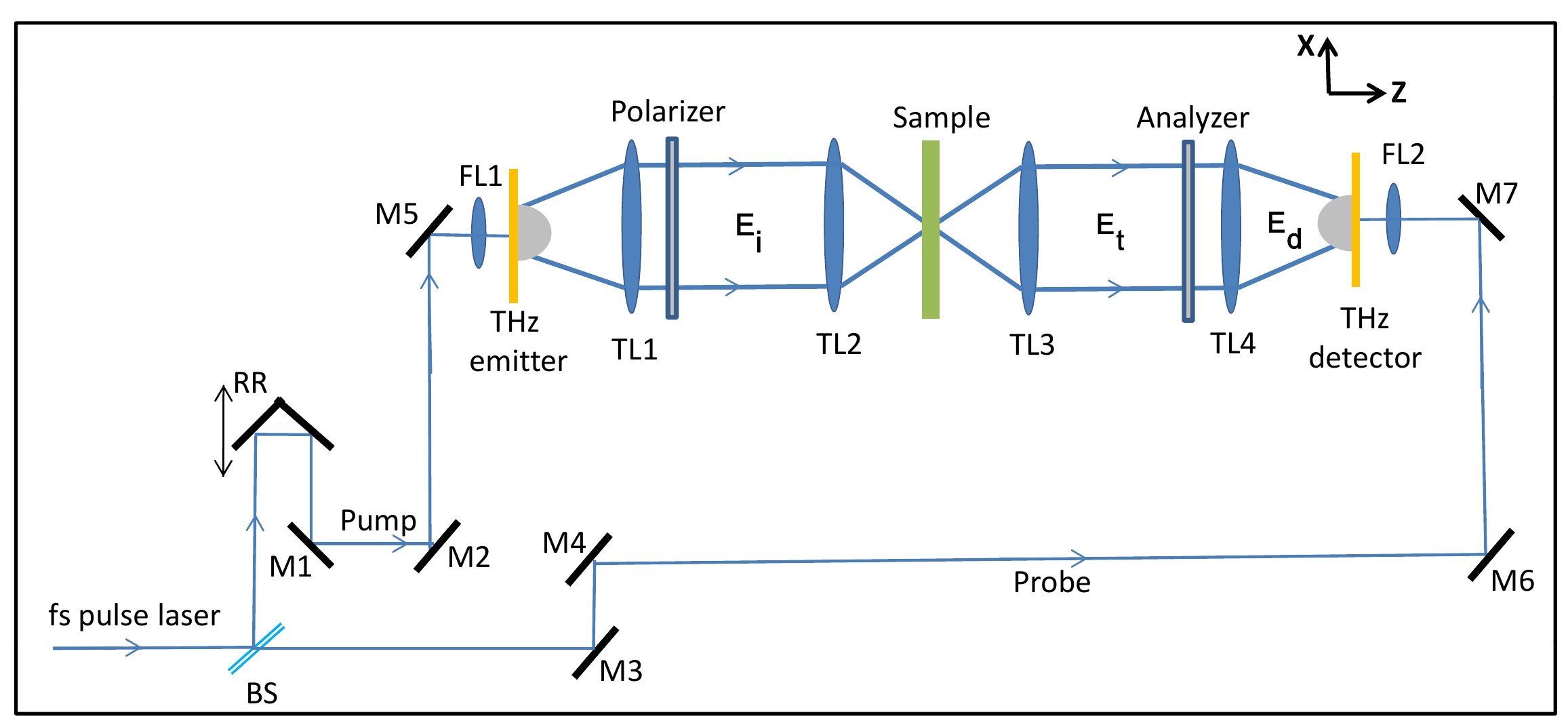}}
\caption{Experimental THz-TDS setup for polarization state measurement (M:mirror, RR: retro-reflector, BS: optical beam splitter, FL: optical focusing lens, TL: THz lens).}\label{Fig1}
\end{figure*}

Fig. \ref{Fig1} illustrates our terahertz time-domain polarization spectroscopy setup.   It uses an $8f$ confocal geometry with THz TPX lenses, which are less prone to misalignments and polarization distortion as compared to off-axis parabolic mirrors. Two identical photoconductive antennas with substrate lens are used as emitter and detectors. A rotatable analyzing polarizer is placed in the collimated beam immediately before the detector, and a fixed polarizer is placed immediately after the emitter. Polarizers were wire grid with wire diameter and spacing of $10~\mu$m and $25~\mu$m, respectively, and field extinction ratio of $\sim$40:1 at 1 THz. The space with terahertz wave propagation is enclosed and purged with dry air during measurements. The laser source is a 800 nm Ti:sapphire femtosecond laser with pulse duration of $<20$ fs and 85 MHz repetition rate, which is divided into pump and probe beams.  The temporal THz pulse is recorded by scanning the retro-reflector and varying the time delay between terahertz pulse and the sampling probe laser. The temporal signal is then taken into the frequency domain through a Fourier transform. 

We model the polarization response of the detector with a polarization vector $\mathbf{P}_d=[p_x~~p_y]$ which is generally frequency dependent. In the frequency domain, the detected signal can be expressed by an inner product as 
\begin{equation}
I(\omega, \phi)=\mathbf{P}_d(\omega)\cdotp\mathbf{E}_d(\omega, \phi),\label{Eq1}
\end{equation} 
where $I$ is the Fourier transform of the detected signal, $\mathbf{E}_d$ is the electric field vector impinging on the detector, $\omega$ is the angular frequency and $\phi$ is the angular orientation of the analyzing polarizer.  The polarization vector can be expressed in the form of a normalized Jones vector as
\begin{equation}
\mathbf{P}_d=\big[\begin{array}{c}
\sqrt{\frac{1+s}{2}}~~~~\sqrt{\frac{1-s}{2}}\exp(j\delta)
\end{array}\big],
\end{equation} 
where the polarization parameters $-1\le s(\omega) \le 1$ and $\delta(\omega)$ are determined through the calibration scheme over the desire frequency range. In Fig. \ref{Fig1}, $\mathbf{E}_t$ represents the transmitted polarization state, and is related to $\mathbf{E}_d$ through the Jones matrix of the rotated analyzing polarizer as
\begin{align}
\Big[\begin{array}{c}
E_x(\omega,\phi)\\E_y(\omega,\phi)
\end{array}\Big]_d=&
\Big[\begin{array}{cc}
\cos^2\phi&\cos\phi\sin\phi\\\cos\phi\sin\phi&\sin^2\phi
\end{array}\Big]\times\nonumber\\
&\Big[\begin{array}{c}
E_x(\omega)\\E_y(\omega)
\end{array}\Big]_t.\label{Eq2}
\end{align} 

In the proposed calibration scheme, the sample is removed and the polarizer at the emitter side is adjusted to yield a known polarization state such that in the absence of the sample $\mathbf{E}_i=\mathbf{E}_t=[1~~0]$. Once the polarization state of $\mathbf{E}_t$ is known, $\mathbf{P}_d$ is determined by solving (\ref{Eq1}) and (\ref{Eq2}) for two scans corresponding to two different angular orientation of the analyzing polarizer. For $\phi=\pm 45^\circ$, $s$ and $\delta$ are readily calculated as
\begin{align}
s&=\frac{1-|r|^2}{1+|r|^2},~
\delta=\arg(r),\nonumber\\
r&=\frac{I(\omega,+45^\circ)-I(\omega,-45^\circ)}{I(\omega,+45^\circ)+I(\omega,-45^\circ)}.
\end{align}

Once the detector polarization vector is characterized, a sample can be put in place, that changes $\mathbf{E}_t$.   Then the new polarization state $\mathbf{E}_t$ is obtained from (\ref{Eq1}) and (\ref{Eq2}) for two angles $\phi=\pm 45^\circ$ as 
\begin{align}
\Big[\begin{array}{c}
E_x(\omega)\\E_y(\omega)
\end{array}\Big]_t=&\frac{1}{p_x^2-p_y^2}
\Big[\begin{array}{cc}
p_x-p_y&p_x+p_y\\p_x-p_y&-(p_x+p_y)
\end{array}\Big]\times\nonumber\\
&\Big[\begin{array}{c}
I(\omega,+45)\\I(\omega,-45)
\end{array}\Big].\label{Eq5}
\end{align} 

In principle any two orientation of the analyzing polarizer can be chosen for this analysis as long as they satisfy the following conditions; $\phi_1,~\phi_2\not=\pm 90^\circ$ and $|\phi_1-\phi_2|\not= 0^\circ,\pm 180^\circ$, however, it has been shown that $\phi_1=+45^\circ$ and $\phi_2=-45^\circ$ provides the best accuracy \cite{Dong_Oct2010}.

\begin{figure}[htb]
\centerline{\includegraphics[width=7.5cm]{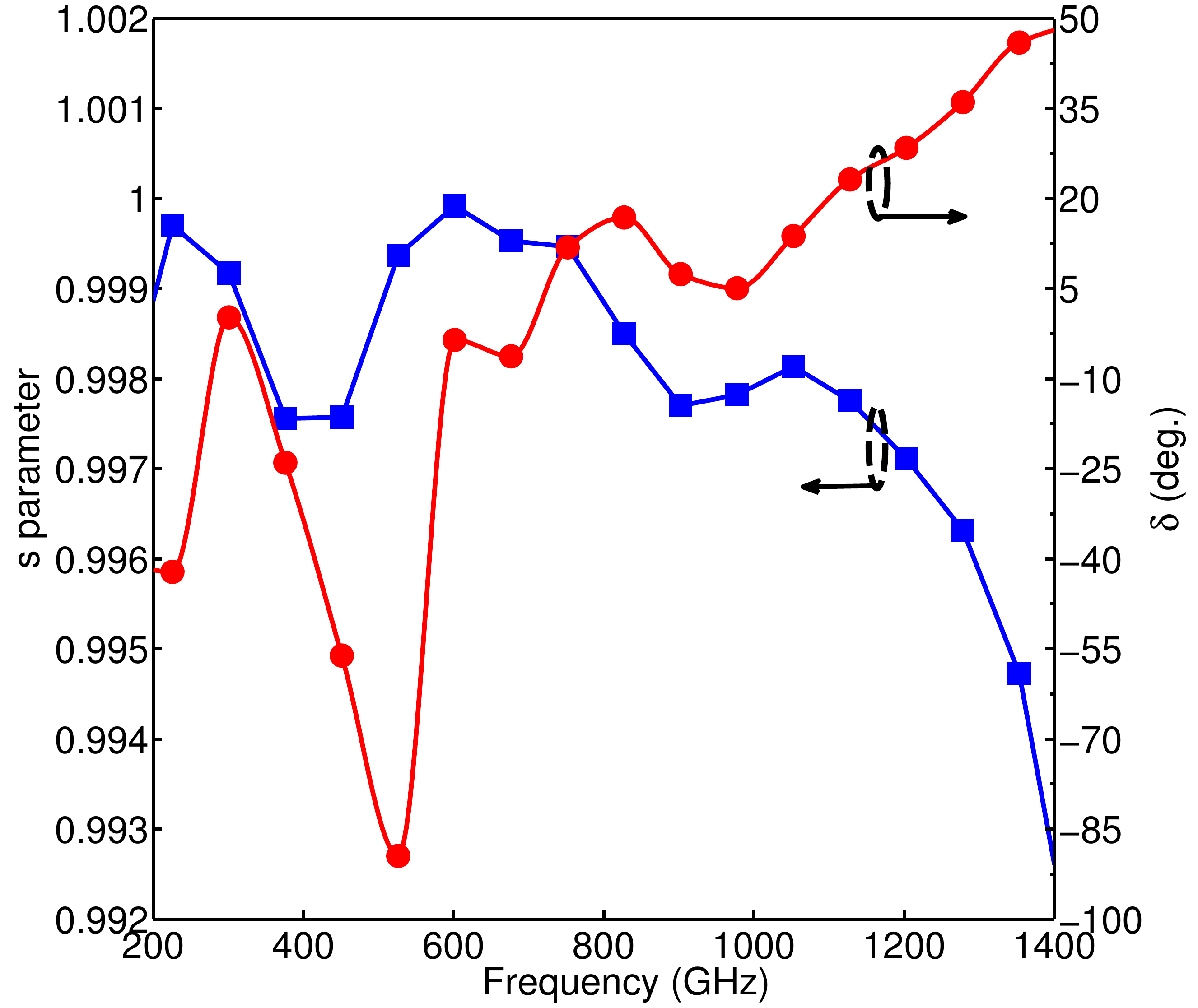}}
\caption{Measured polarization parameters $(s,\delta)$ of  a 20 $\mu$m dipole with 5 $\mu$m gap size.}\label{Fig2}
\end{figure}

Two commercially available photoconductive dipole antennas (TERA8-1) with 20 $\mu$m length and 5 $\mu$m gap sizes from MenloSystems were used.  Fig. \ref{Fig2} shows the measured $(s,\delta)$ parameters over the frequency range for one of these dipoles. For an ideal detector with linear polarization along the x-axis $(s,\delta)$ would equal $ (1,0)$ ($\mathbf{P}_d=[1~~0]$) for all frequencies.   It is clear that errors introduced from performing polarimetry without a calibration step, while small, are not negligible.

\begin{figure*}[htb]
\centerline{\includegraphics[width=10cm]{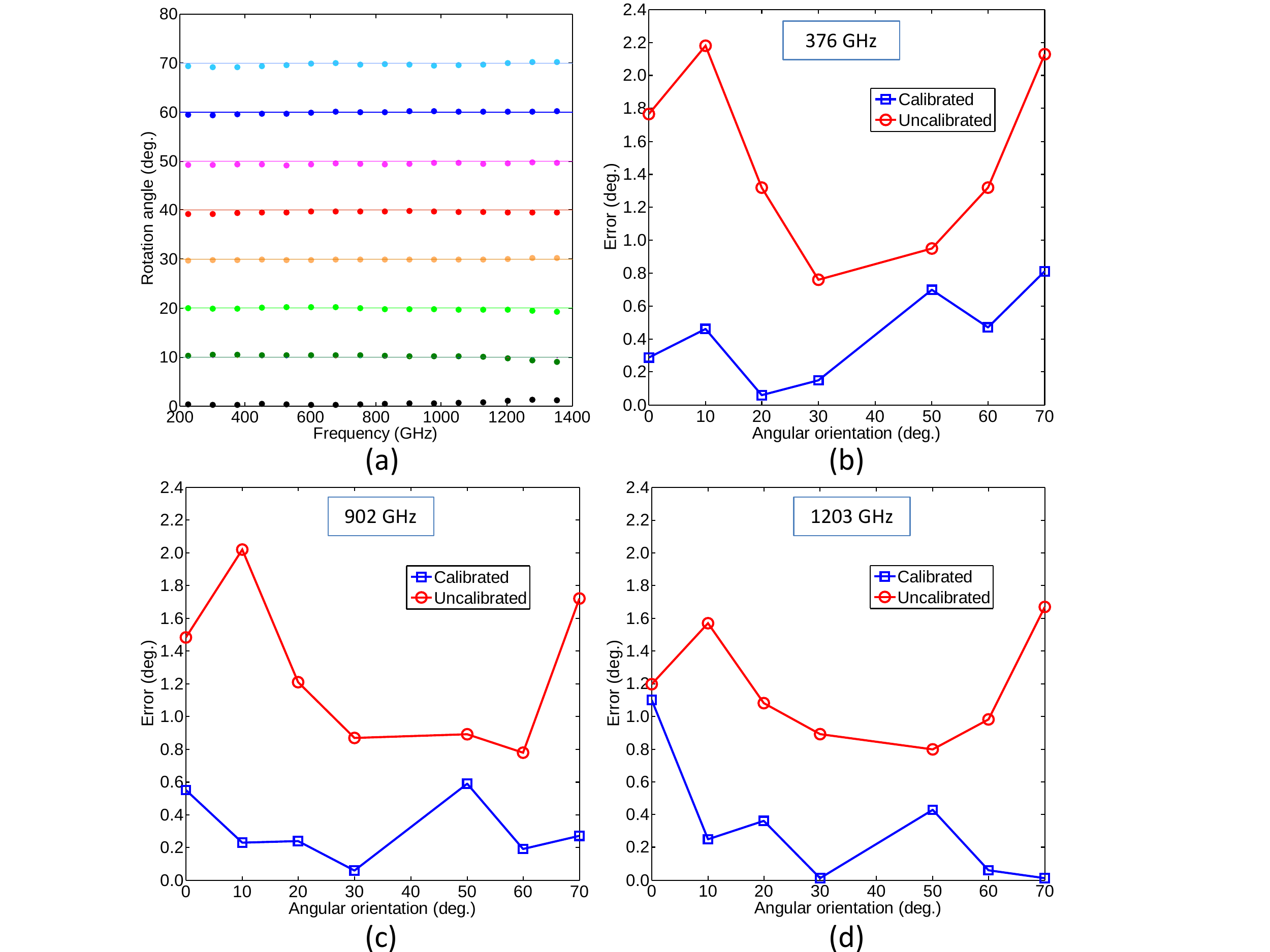}}
\caption{(a) Extracted rotation angle of the sample polarizer, (b)-(d) comparison of the error of the extracted rotation angle between calibrated and uncalibrated measurements for displayed frequencies.}\label{Fig3}
\end{figure*}

In order to evaluate the improvement from calibrating out the detector response, a wire grid polarizer similar to the analyzing polarizer was used as a sample. In this case, the polarization state after the sample polarizer is well known and can be compared with that from measurement. The sample polarizer was installed in a precision rotation stage with $0.08^\circ$ accuracy. The polarizer axis was determined accurately by the diffracted pattern of a red laser passing through the wire grid. The sample polarizer was rotated from $0^\circ$ (polarizer axis along x-axis) to $70^\circ$ with $10^\circ$ increment.  Polarization states were measured for each position of the sample polarizer by using (\ref{Eq5}). Fig. \ref{Fig3}(a) shows the extracted angle from the measured polarization state. An excellent agreement between set and extracted rotation angle is shown in Fig. \ref{Fig3}(a).  Figs. \ref{Fig3}(b)-(d) compare the error of the extracted rotation angle between calibrated and uncalibrated measurement for displayed frequencies. Based on this comparison the root-mean-square (RMS) error is reduced considerably by at least $60\%$ after applying the calibration.

\begin{figure}[htb]
\centerline{\includegraphics[width=7.5cm]{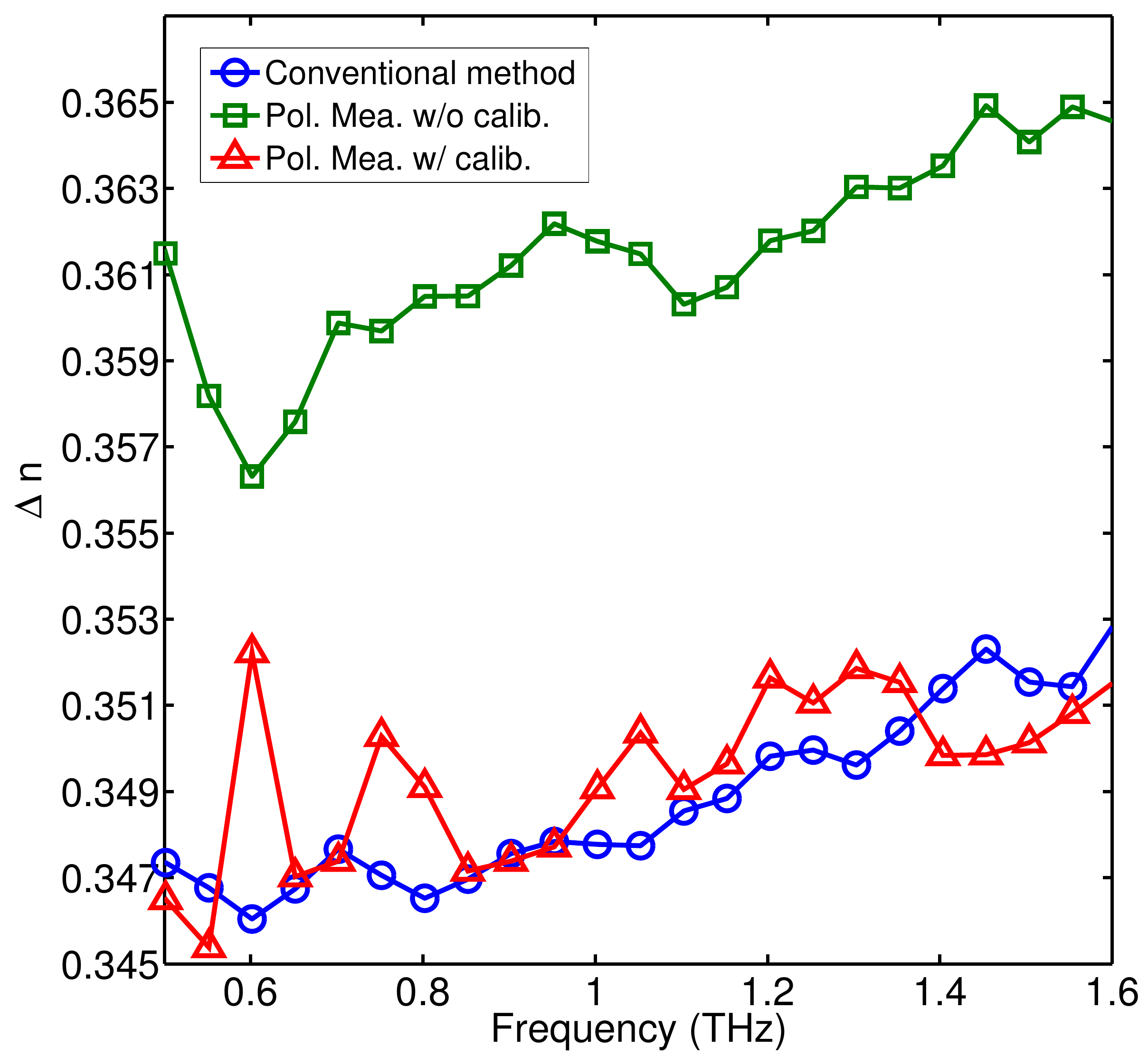}}
\caption{Measured birefringence of a sapphire wafer obtained from polarimetry with and without calibration, and from conventional method.}\label{Fig4}
\end{figure}

As a proof-of-principle for this method of obtaining polarization states, we studied the effect of the proposed calibration scheme on measuring birefringence through polarimetry. A 50 mm-diameter sapphire wafer with 0.47 mm thickness and C axis in the plane of the wafer was used as sample. For a uniaxial crystal with its optical axis in the x-y plane, the phase retardation ($\Delta$) between the optical axis and its perpendicular direction and the angular orientation ($\theta$) of the optical axis itself can be obtained simultaneously from its Jones matrix as
\begin{align}
\theta&=\frac{1}{2}\tan^{-1}\left[\frac{1}{\Re(R)}\right],~\Delta=-2\tan^{-1}\left[\frac{1}{\sin(2\theta)\Im(R)}\right],\nonumber\\
R&=\frac{E_{xt}(\omega)}{E_{yt}(\omega)},\label{Eq6}
\end{align} 
where $\Re(.)$ and $\Im(.)$ denote the real and imaginary operators, respectively, and $E_{xt}$ and $E_{yt}$ are the x- and y-component of the transmitted polarization state, respectively, when $\mathbf{E}_i=[1,0]$. Fig. \ref{Fig4} compares the measured birefringence magnitude of the sapphire wafer, extracted from the phase retardation ($\Delta$) in (\ref{Eq6}), with and without applying the calibration scheme. For this measurement, the angular orientation $\theta$ was obtained from (\ref{Eq6}) as $\theta=43.5^\circ$ by averaging $\Re(R)$ over the frequency range where it behaves smoothly. Also shown in Fig. \ref{Fig4} is the birefringence calculated from the difference between the indices of refraction for ordinary and extraordinary rays ($\Delta n=n_e-n_o$) using a more conventional method of aligning the THz electric field along crystalline axes in seperate measurements.  As seen in Fig. \ref{Fig4}, the calibration has considerably improved the birefringence extracted from polarization measurement when compared to the uncalibrated measurement. It should be emphasized that in the polarization measurement, unlike the more conventional method, it is not necessary to know $\theta$  $a$ $priori$.

This work was made possible by support from DARPA YFA N66001-10-1-4017 and the Gordon and Betty Moore Foundation.


\pagebreak

\section*{Informational Fourth Page}


\end{document}